\documentclass[12pt]{article}
\usepackage{geometry} 
\usepackage{epstopdf}
\usepackage{epsfig}
\usepackage{amsmath}
\usepackage{booktabs}
\usepackage{subfig}
\usepackage{natbib,hyperref}         
\title{How much diversification potential is there in a single market?
\\ Evidence from the Australian Stock Exchange }

\author{Libin Yang$^{1}$,   William Rea$^{1}$,  and Alethea Rea$^{2}$, \\
1. Department of Economics and Finance, University of Canterbury, \\
New Zealand \\
2. Data Analysis Australia, Perth, Australia }

\begin{document}

\maketitle

\begin{abstract}
We present four methods of assessing the diversification potential within
a stock market, two of these are based on principal component analysis.
They were applied to the Australian stock exchange for the years 2000 to
2014 and all show a consistent picture. The potential for diversification
declined almost monotonically in the three years prior to the
2008 financial crisis. On one of the measures the diversification potential 
declined even further in the 2011 European debt crisis and the
American credit downgrade. 
\end{abstract}

\begin{description}
\item[Keywords: ] Principal component analysis, stock selection, 
diversification, stock portfolios
\item[JEL Codes: ] G11
\end{description}

\section{Introduction}\label{sec:Introduction}

The purpose of diversification in a portfolio is widely understood 
from both a practical and a theoretical viewpoint. For example,
\citet[][Ch7]{Bodie2011}, in common with many other authors,
divide the total risk 
of an asset into two parts, the idiosyncratic risk and the
systematic risk. Diversification is the grouping together of
different assets in order to  reduce the
idiosyncratic or diversifiable risk, leaving a portfolio with, in the
ideal case,
only the systematic risk, see \cite{Bodie2011} Figure (7.1). However,
\cite{meucci2009} points out ``\dots there exists no broadly accepted,
unique, satisfactory methodology to precisely quantify and manage
diversification.''

In addition to being unable to precisely quantify diversification
potential, there is the problem facing investors that the level of 
diversifiable risk,
that is the risk that can be eliminated through diversification,
does not appear to be constant but is time varying 
both within and between markets.
A consequence of this problem is that even if the assets
held within a portfolio are not changed the amount of diversification
that portfolio has rises and falls with changing market conditions.

Thus there is need for tools to help investors understand how much
potential for diversification exists to enable them to manage their
portfolios effectively. A significant body of literature relavant to this
topic already exists in two forms. One tries to directly assess the
diversification potential which exists in a market or set of markets.
For example, \cite{DeFusco1996} use co-integration methods to 
investigate the diversification potential in 13 emerging capital markets.
The other is concerned with the estimation of systemic risk (not to be
confused with systematic risk), see, for example,
\cite{Kritzman2011}, \cite{Billio2012}, and \cite{Zheng2012}.
 Much of the literature in this second group uses
principal component analysis (PCA) for some of the work of estimation.
(See \cite{Jolliffe1986} for a detailed description of PCA.)
 
PCA is a standard method in statistics
for extracting an ordered set of uncorrelated
sources of variation within a multivariate system. Given that financial
markets are typically characterised by a high degree of multicollinearity,
implying that there are only a few independent sources of information 
in a market, PCA is an attractive method to apply.

Looking at the results of a PCA from a theoretical point of view
\cite{kim2005} decomposed the correlation matrix of a selection of 135 
stocks which traded on the New York Stock Exchange into three parts based on 
the Spectral Decomposition Theorem\footnote{More information of Spectral 
Decomposition Theorem, see \citet[][p13]{Jolliffe1986}.}. 
They assigned the following meaning to the  principal components:
\begin{enumerate}
\item The first principal component (PC1) with the largest eigenvalue 
represents a 
market wide effect that influences all stocks. In the financial
literature this is often called the systematic risk.
\item A variable number of principal components (PCs) following 
the market component which
represent synchronized fluctuations associated with specific groups of stocks.
\item The remaining PCs indicate randomness in the price
fluctuations (noise). There were believed to contain no useful financial 
information and hence were eliminated from further investigation.
\end{enumerate}

Systemic risk is defined as ``the risk associated with the whole 
financial system, as 
opposed to any individual entity or component''. It can also
be defined as ``any set 
of circumstance that threatens the stability of financial system, and so 
potentially initiates financial crisis" \citep{Zheng2012}. 
For an investor seeking to manage a portfolio, rather than a 
regulator seeking to ensure financial system stablility,
systemic risk can be thought of as the ratio of systematic risk to 
idiosyncratic risk. An increase in the systemic risk suggests that the 
systematic risk as a proportion of the total risk increases. Put another way,
if the systemic risk increases, the 
amount of idiosyncratic risk, the diversifiable risk, as
a proportion of total risk decreases leaving
the investor poorly prepared for any negative shocks to the financial
markets. 

After the financial crisis in 2008 literature relating to systemic risk 
has grown. There have been three groups of empirical studies on 
systemic risk. 
\begin{enumerate}
\item Literature focused on contagion, spillover effects and 
joint crashes in financial markets \citep{adrian2007, wang2011, 
Kritzman2011, Billio2012}. Those studies were based on the analysis of 
interconnectedness among market security returns and our analysis 
below follows in their steps.
\item Empirical studies on systemic risk which include research 
on the auto-correlation in the number of bank defaults, bank returns, and 
fund withdrawals \citep{de2000, lehar2005, kenett2012} 
\item Research on bank 
capital ratios and bank liabilities \citep{aguirre2004,bhansali2008,brana2009}.
\end{enumerate}

When the market becomes more connected, that is, the correlations
between assets rises, the systemic risk is higher in 
the sense that shocks propagate more quickly and broadly. For 
this reason, monitoring the time evolution of correlation is critical
in portfolio management. 

Other research has shown that the security correlations change in different 
time periods. \cite{butler2001} and \cite{campbell2002} reported that the 
market correlation increased in bear markets. \cite{ferreira2004}, 
\cite{hong2007}, and \cite{cappiello2006} reached the same conclusions for 
global industry returns, individual stock returns and international bond 
returns.

Instead of comparing different time periods, many recent papers have 
applied PCA 
to investigate correlation using a sliding window approach. 
\cite{fenn2011} applied PCA to study the evolution of correlation in a 
diverse range of asset classes.
They asserted that increases in the variance explained by 
PC1 implied that there was more common 
variation in financial 
markets. Moreover, they emphasized that the variance explained by PC1
might be the result of either (1) increases in the correlations 
among a few 
assets or (2) increases in market-wide correlation. 
The first case will have less impact 
on the ability to diversify because an investor could simply move 
investments to less 
correlated assets. In contrast, it becomes much more difficult to reduce 
risk by diversifying across different assets if there is a market-wide 
correlation increase. For example, they reported that the sharp increase
in variance explained by PC1 
on 15 Sep 2008 was the case of a market-wide correlation increase precipitated
by Lehman Brothers filing for 
bankruptcy and Merrill Lynch agreeing to be taken over by the Bank of America.

\cite{Kritzman2011} introduced a measure of systemic risk called the 
absorption ratio. If differs from the measure used by \cite{fenn2011}
in that it is the fraction of variance absorbed by a fixed, finite 
number of PCs rather than PC1 alone. They 
reported that most financial 
crises were coincident with positive shifts of the absorption ratio. These 
crises include the Asian Financial Crisis in 1997, Russian default and 
LTCM collapse in 1998, the housing bubble in mid-2006, 
and the Lehman Brothers default 
in 2008. Another interesting finding in this paper is, in most cases, stock 
prices changed significantly when the absorption ratio reached its highest or 
lowest level.

\cite{Zheng2012} not only looked at the absolute value of variance explained 
by PC1, they also computed the change in the variance 
explained to capture the systemic risk. They obtained similar findings to 
\cite{Kritzman2011} that both the absolute value and change of variance 
explained by PC1 increased during a financial crisis. 
But they reported that the moving window size and the time length used to 
calculate the change had an impact on the date of the spike. The spike of 
absolute value of variance explained by PC1 occurred 
later when the moving window size was larger and saturated after approximately 
20-month time window. 

In this paper we confine ourselves to the
Australian stock market and present several related tools which can
be used the gauge the diversification potential in that market. However, the 
methods presented are quite general and can used to assess the
diversification potential across multiple asset classes.

The structure of the paper is as follows. Section (\ref{Sec:DataDescript})
describes the data and methods, Section (\ref{sec:Results}) contains
the results,
Section (\ref{sec:Discussion}) 
contains the discussion and conclusions.

\section{Data and Methods}\label{Sec:DataDescript} 

This section is structured as follows; Section (\ref{sub:Data}) describes
the data we obtained and the preparation of the return series,
Sections (\ref{sub:KMO}) through (\ref{sub:Diversification}) then describe
the four methods of analysis which we applied to the return series. All
four methods rely on analysing either a correlation or a covariance
matrix, the generation
of these matrices is described in Section (\ref{sub:KMO}).
The first two methods, in
Sections (\ref{sub:KMO}) and (\ref{sub:PComponents}), give us some
insight into the connectedness of the market  from which we
can infer the diversification potential which is there. The
second two, in Sections  (\ref{sub:PCAStock}) and (\ref{sub:Diversification}) 
give us a more direct measure of diversification potential.

A rolling window approach was applied in our estimation process. We
performed each analysis on a window size of two years (equivalent to 504 trading
days) at weekly intervals. This resulted in 602 data points for each 
of our four analysis methods presented below.

\subsection{Data}\label{sub:Data}
Our research is based on the Australian market. The main index
for the market is the ASX200, which
is a market capitalization weighted index of the 200 largest 
shares by capitalization listed on the Australian Securities Exchange. 
The index in its current form was created on 31 March 2000. 
We investigated the 
constituents of the ASX200 index from inception to February 2014. 
The ASX200 index is a capitalization index and so
does not adjust for dividends.  In our
research we calculated the returns for
all constituents which included the dividends paid.

There was a high frequency of stocks that were added to or deleted from the 
index from time to time, so we identified all stocks which had been in the 
ASX200 for the whole study period. After adjusting for mergers, acquisitions, 
and name changes we obtained a final data set of 524 unique stocks. 
We obtained daily closing prices and dividends for each stock from the 
SIRCA database\footnote{\url{http://www.sirca.org.au/}}. 
All the prices and dividends were adjusted to be based on the AUD. 
The return series was calculated from the price and dividend data, see 
Appendix \ref{App:Return} for details.

We extracted a set of stocks that had complete return information for the 
whole study period, and there were 156 such stocks. The remaining 368 
stocks were either listed after April 2000 or delisted before February 2014. 

\subsection{Kaiser-Meyer-Olkin Test}\label{sub:KMO}

Correlation and covariance matrices were generated from the
return series with the {\tt cor} and
{\tt cov} functions respectively in
the {\tt stat} package in 
base {\tt R} \citep{R} 
on a rolling window of 504 trading days, which is equivalent to
two calendar years. The correlation matrices were for use
with the KMO test described in this section and the two tests
involving PCA.
The covariance matrices were for use with the 
diversification ratio described in
Section (\ref{sub:Diversification}) below.

The Kaiser-Meyer-Olkin (KMO) measure of sampling adequacy 
\citep{kaiser1970,kaiser1974} is calculated as 
$$
\text{KMO}=\frac{\sum\sum_{j\ne k}r_{jk}^2}{\sum\sum_{j\ne k}r_{jk}^2+\sum\sum_{j\ne k}
q_{jk}^2}
$$
where the $r_{jk}$ are the original off-diagonal correlations and
the $q_{jk}$ are the off-diagonal elements of the partial-correlation
matrix. Thus the KMO statistic is a measure of how small the partial 
correlations are, relative to the original  correlations, the smaller
$\sum\sum_{j\ne k} q_{jk}^2$ are, the closer the KMO statistic will be to one.
A KMO value of $0.5$
is the smallest KMO value that is considered acceptable for a PCA.

We calculated the KMO statistic in rolling windows of different sizes for the 
$156$ stocks which had complete data and settled on a window
size of two years or $504$ trading days as indicated above because this
always gave a KMO value greater than 0.5. 

The KMO test was performed using functions in the {\tt R} package
{\tt psych} \citep{psych}.

\subsection{Principal Component Analysis}\label{sub:PComponents}

PCA can be applied to either a correlation matrix or a covariance matrix.
All PCAs reported in this paper were carried out on correlation matrices.

PCAs were carried out using the function 
{\tt eigen} in base
{\tt R} on a rolling window of 504 trading days. For a correlation
matrix the total variation is equal to the number of variables
in the matrix. Thus for our matrices this was 156. To obtain the
percentage of variance explained by PC1 if $E_1$ is the eigenvalue of PC1
then
$$
\text{Percent Variance Explained}=\frac{E_1}{156}\times 100.
$$

\subsection{PCA Stock Selection}\label{sub:PCAStock}

\cite{Yang2015a} presented a method for stock selection using principal
component analysis. Their procedure is as follows:
 \begin{enumerate}
 \item Apply PCA to the correlation matrix of a stock market.
 \item Associate one stock with the highest coefficient in absolute value
with each of the last $m$ principal components that have eigenvalues less
than a certain level called the deletion criteria, then delete
those $m$ stocks.
\item A second or subsequent PCA is performed on retained stocks. 
The same procedure described in step 2
is applied to the output of the PCA and, if necessary, further stocks are
deleted.
 \item The procedure is repeated until no further deletions are considered
necessary based on a stopping criteria which is a pre-determined minimum
eigenvalue of the last principal component. 
 \end{enumerate}
We used their proceedure with a deletion criteria of 0.7 and a 
stop criteria of 0.5.

Intuitively, the procedure seeks to step-wise remove the most highly
correlated stocks in the sample leaving the most independent stocks. For example,
if two stocks are highly correlated they are likely to be found
in the same high numbered PC each with a high loading. The procedure
will then elminate the one with highest loading leaving only one of the
original pair in the sample. From a diversification point of view elimininating
one of a highly correlated pair of stocks will result in only a small
loss of diversificaion potential, most of the potential will still
be in the sample in the form of the retained stock.

\subsection{Diversification Ratio}\label{sub:Diversification}

The diversification ratio is a measure of the degree of diversification for 
a long-only portfolio introduced by \cite{choueifaty2008}. The 
diversification ratio for a portfolio is defined as
\begin{equation}
\text{DR}_{\omega\in\Omega} = \frac{\omega'\sigma}{\sqrt[]{\omega'\Sigma\omega}}
\label{eqn:Diversification}
\end{equation}
where 
$\omega$ is the weight vector of the portfolio,
$\Omega$ is the set of investible assets,
$\sigma$ is the vector of asset volatilities measured by their 
respective standard deviations and
$\Sigma$ is the variance-covariance matrix of the returns for the N 
assets. 
The numerator of the diversification ratio is then the weighted average 
volatility of the individual stocks and the denominator is the portfolio 
standard deviation. By this definition, the higher the diversification ratio, 
the better the degree of diversification. If a portfolio is completely 
non-diversified, in the case of single-asset portfolio, the diversification 
will achieve its lower bound of 1. The diversification ratio was
calculated using custom written R code.

\section{Results}\label{sec:Results}

The
results are presented in Figures (\ref{fig:MSA}) through
(\ref{fig:Diversification}) and discussed in turn in the sections
below. They are presented in the order given in the previous 
section. 

\subsection{KMO Test}

We discussed the KMO measure of sampling adequacy 
in Section (\ref{sub:KMO}). It is typically used prior to
running a PCA to assess whether a PCA is worth
performing (this was a serious issue when computing power was
limited). Because in our sample it is a test of
the degree of common variation among stocks it gives us 
a tool to assess the market connectedness and therefore the 
potential for diversification. 

\begin{figure}
\centering
\includegraphics[width=\linewidth]{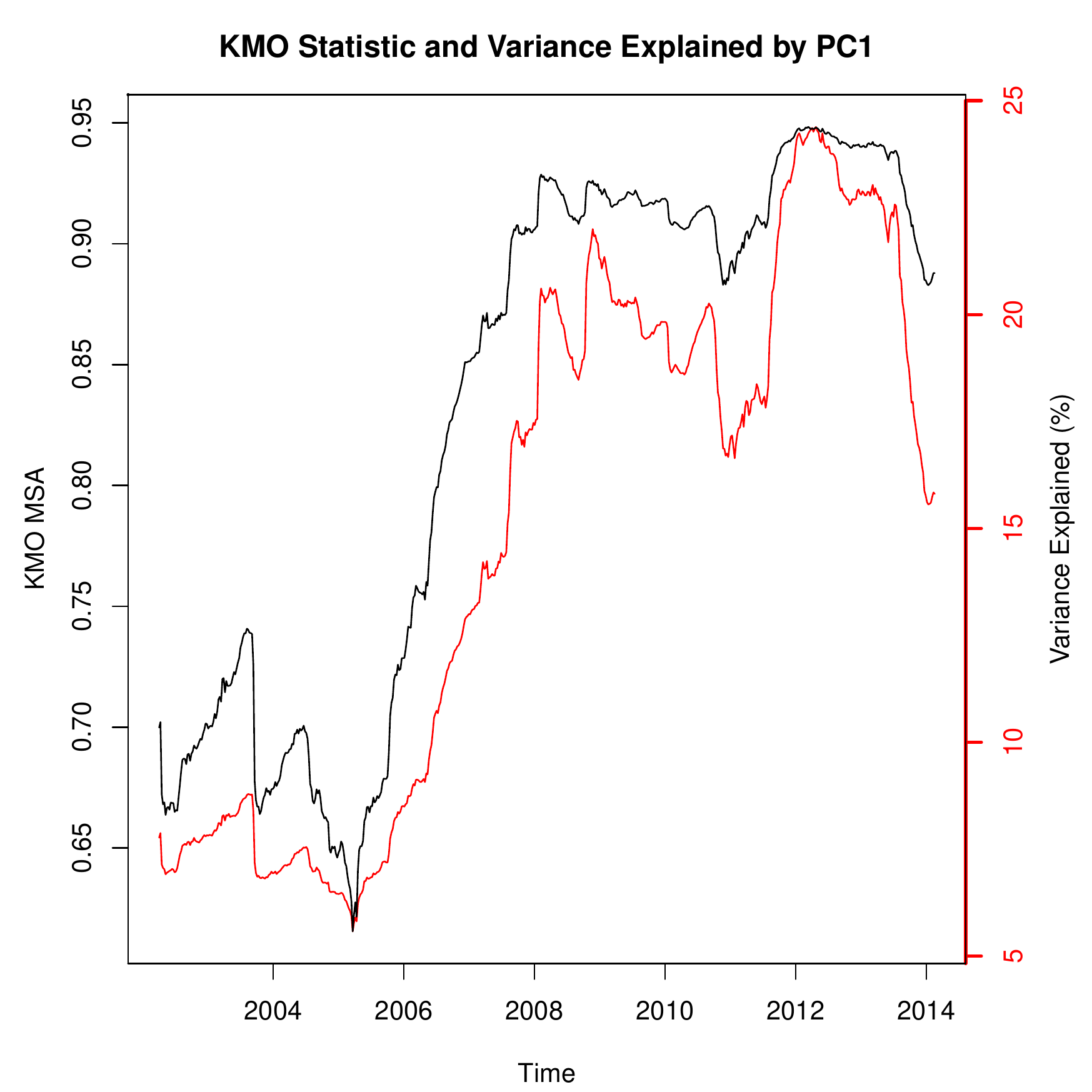}
\caption{KMO measure of sampling adequacy for 156 stocks and the variance 
explained by PC1. Both measures were calculated weekly 
using a rolling window approach with window size two years (504 trading days).}
\label{fig:MSA}
\end{figure}

Figure (\ref{fig:MSA}) presents KMO statistics from 2002 to 2014. For
comparision purposes 
the variance explained by PC1 is plotted on the same
graph. A higher level of the KMO statistic 
indicates more common variation among stocks which suggests less potential 
for diversification. We can see that the KMO statistic and the variance 
explained by PC1 evolved closely over time. 

\subsection{Market wide effects: Principal Component One}\label{ssec:PC1}

As indicated in Section (\ref{sec:Introduction}) other authorities have
used PCA to analyse market connectedness and systemic risk and have
reported that financial markets have become more integrated during market 
crashes.  We follow \cite{fenn2011} and \cite{Zheng2012} and used 
the variance explained by the PC1.

\begin{figure}
\centering
\includegraphics[width=\linewidth]{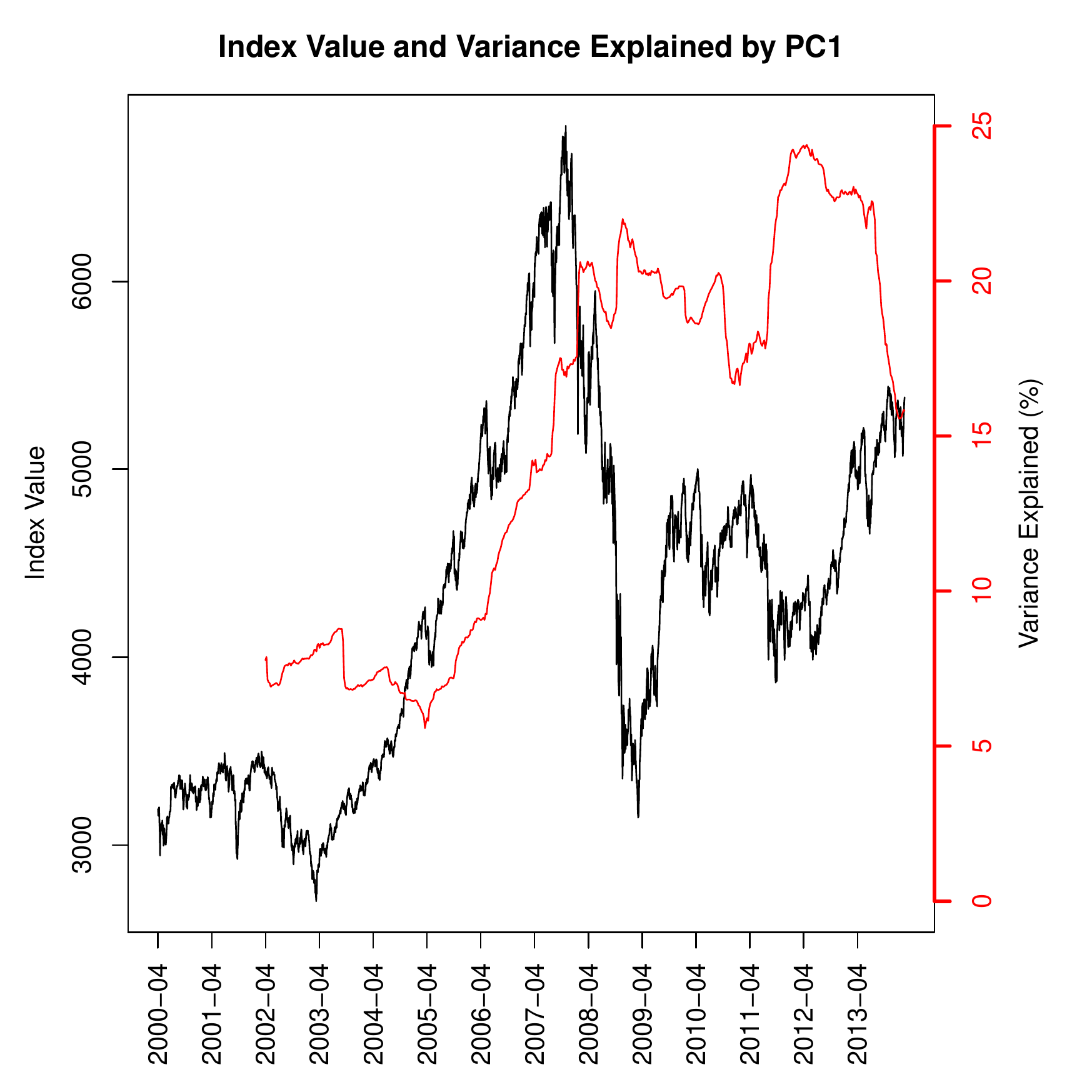}
\caption{Variance explained by PC1 with the ASX200 index value. The plot of
index values includes the first two year period used for the estimation
of the PCA. }
\label{fig:pc1indexvalue}
\end{figure}

\begin{figure}
\centering
\caption{Variance explained by PC1 and the ASX200 index returns. Both the 
variance explained by PC1 and the ASX200 index returns were calculated 
weekly using a rolling window size of two years 
(equivalent to 504 trading days). }
\label{fig:pc1index}
\centering
\includegraphics[width=\linewidth]{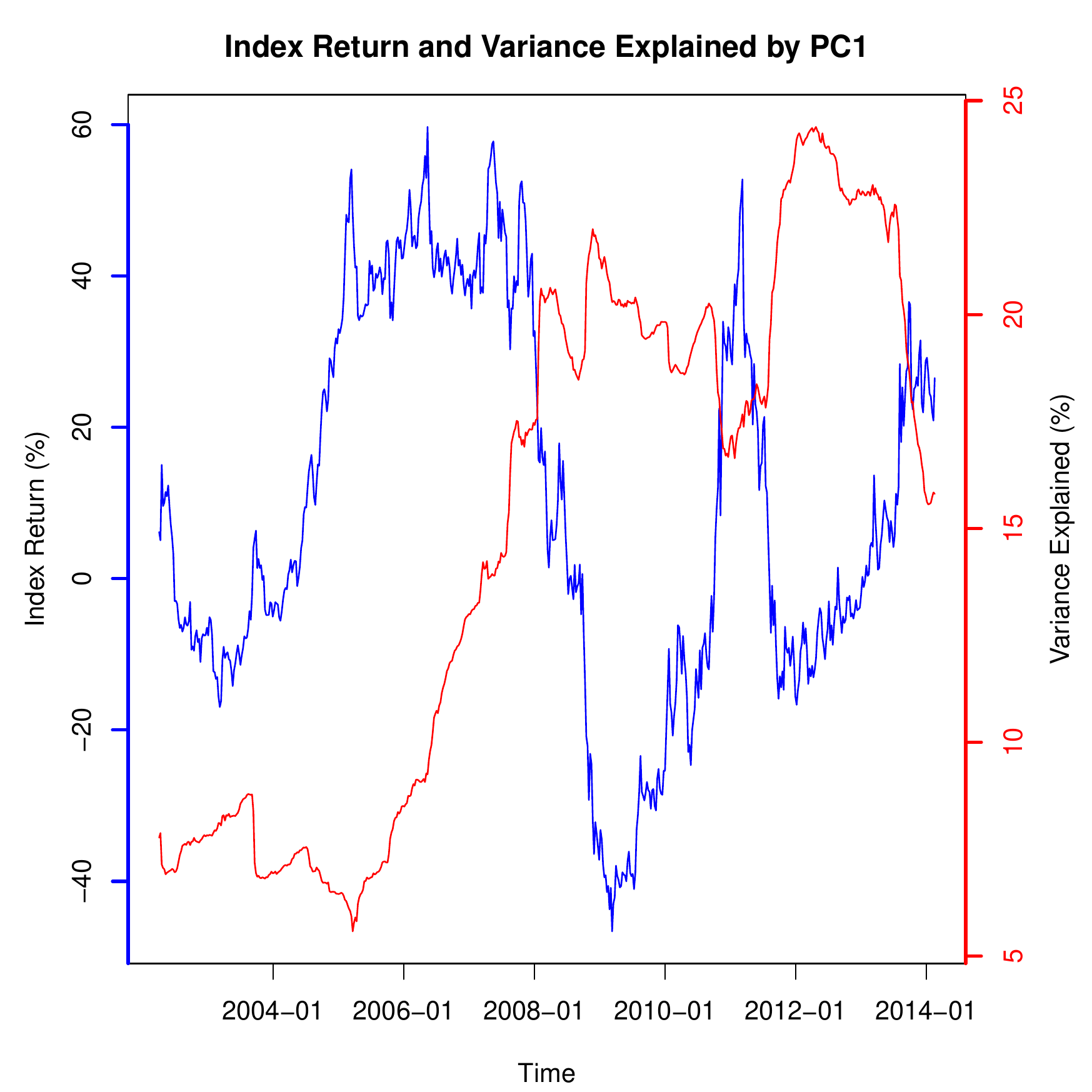}
\label{fig:pc1indexreturn}
\end{figure}



Figures (\ref{fig:MSA}), (\ref{fig:pc1indexvalue}), (\ref{fig:pc1indexreturn}) 
and (\ref{fig:Diversification}) each contain a plot of
the variance explained by PC1. It is paired with the 
KMO statistic, the ASX200 index value, the index returns,
and the diversification ratio respectively. 
Figure (\ref{fig:pc1indexvalue}) shows the index for
the full study period. There are no results for the variance explained
by PC1 in the first two years because that is the estimation period
for the PCA. The value for variance explained has been plotted to
coincide with the end of the estimation period. 
There was an almost five year period from 2003 to 2008 in
which the market steadily rose. The variance explained by PC1 began to
steadily rise once the tail of the rolling window crossed the start of
the rise. The rise was from a low of about five percent to about 20 percent
close to the start of the financial crisis in 2008.

Figure (\ref{fig:pc1index}) presents the same information as
Figure (\ref{fig:pc1indexvalue}) in a somewhat different form. 
The index values have been converted to estimation period returns.
The index returns stayed high when the variance explained increased 
steadily from 
2005 to 2008. Once the variance explained reach the first peak, the index 
returns started to drop significantly. The market remained tightly integrated 
until the beginning of 2010, even when index returns started to recover from 
the beginning of 2009. This suggests that the Australian stock market was 
still extremely fragile and therefore vulnerable to large shocks during
this period. 

There was a small drop of variance explained at the end of 2010 and the 
index returns reached its second peak. The variance explained stared to rise 
again at the end of 2011 and reached highest point during the study period 
at beginning of 2012. The reason for this increase of variance explained 
appears to be because of market-wide increases in correlation precipitated
by the European sovereign debt crisis, and fears over the global economy. 
At the end of our sample period, the variance explained by principal 
component one had decreased significantly and the index returns had 
partially recovered from the drop in late 2011.

\subsection{Number of Stocks Required for a Diversified Portfolio}

The stock selection procedure described in Section (\ref{sub:PCAStock})
systematically selects a subset of stocks which are intended to
maximise the diversification potential of the subset based on the 
correlations between them. 
However, the correlations between the
stocks have changed over time and this affected the number 
of stocks selected. 
We hypothesised that during the periods 
of a more connected market there would be less risk sources. This means that
without altering the selection criteria
one should expect a smaller number of stocks being selected to describe the 
market. This is indeed what we saw.

\begin{figure}
\centering 
\caption{The number of stocks selected by PCA over time. A stocks selection 
procedure of 0.7 deletion criteria and a 0.5 stop criteria was used. 
The selection procedure was applied on a rolling window basis with window 
size of two years (504 trading days). }
\label{fig:nostocks0705}
\includegraphics[width=0.8\linewidth]{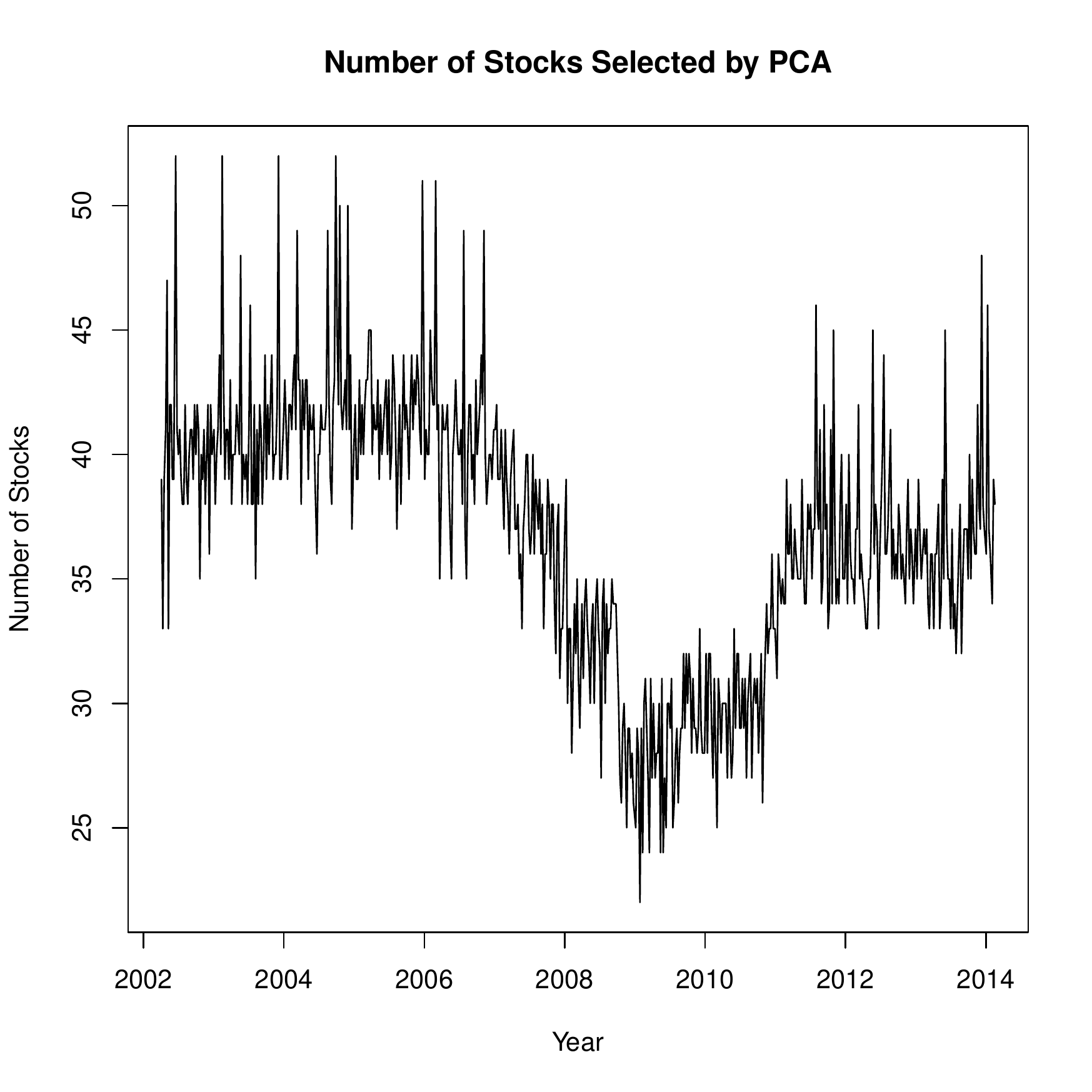}
\end{figure}

In Figure (\ref{fig:nostocks0705}) presents the number of stocks
selected. The results show a lot of fluctuation in the number of
stocks selected. Nevertheless, consistent with the other
results presented in this paper,
the number of stocks selected decreased starting in 2006 and reached their 
lowest level
in late 2009. The market had already become concentrated and offered fewer 
diversification opportunities before the 2008 financial crisis started.
When the market became less tightly coupled, the number of stocks increased
again. 

This trajectory of number of selected stocks moved in an opposite way to 
the variance explained by PC1 (see Figure \ref{fig:pc1indexvalue}). 
This illustrates that the number of stocks needed to diversify a portfolio 
is not constant through time. With the number of major stock market risk 
sources changing, a portfolio can be considered diversified consistently 
only if it is adaptive to the change. In other words, the number of stocks 
included to diversify major risk sources should change based on the number 
of risk sources in the market. Thus, a portfolio that holds the same number 
of stocks or same constituents can only be the best combination to create 
a diversified portfolio at a single point of time. Holding more stocks 
than necessary when the number of major risk sources decreases is 
redundant. On the other hand, holding fewer stocks than required when 
the number of major risk sources increases means that the portfolio is 
under-diversified.

\subsection{Diversification Ratio}

\begin{figure}
\centering
\label{fig:pc1diversificationratio}
\includegraphics[width=\linewidth]{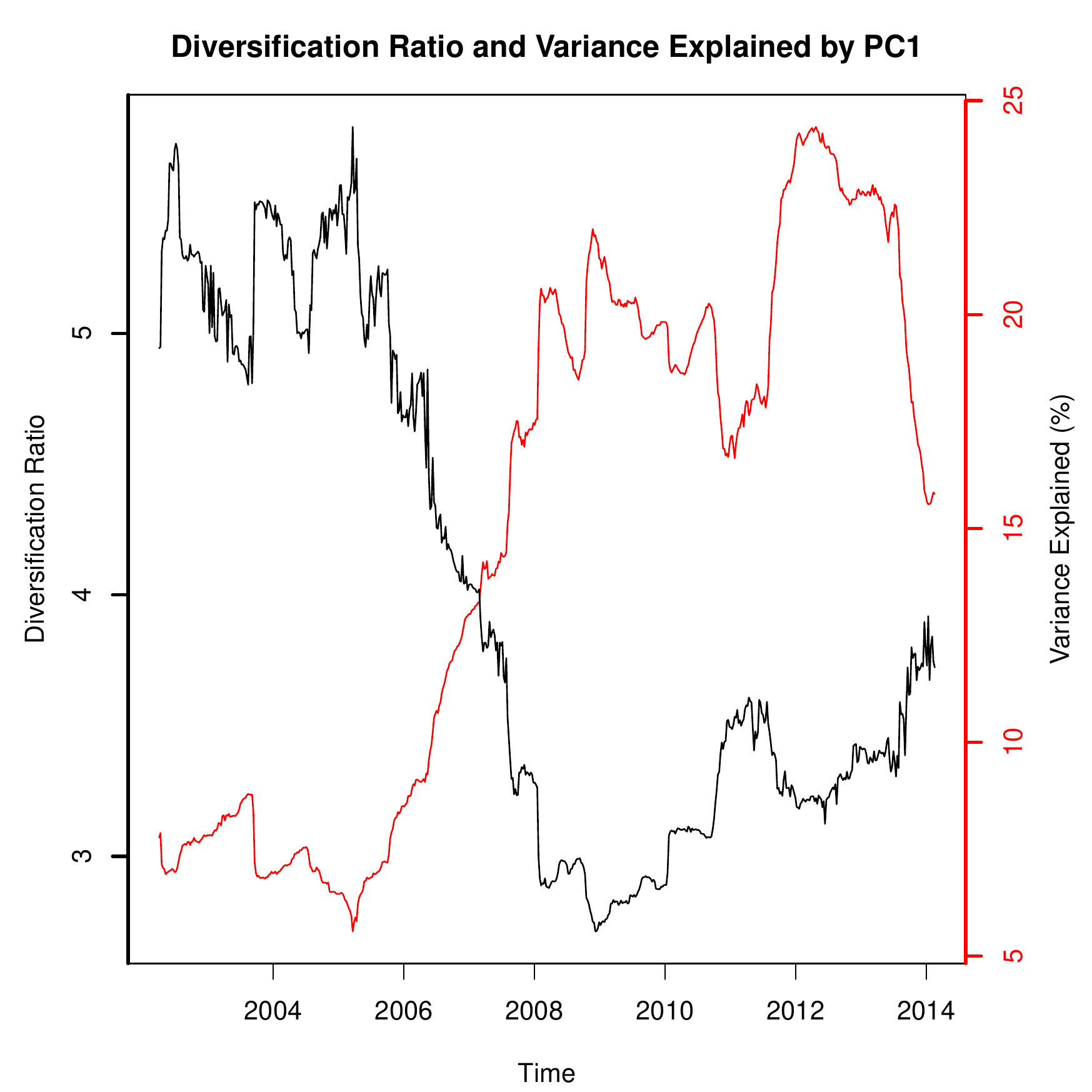}
\caption{Variance explained by PC1 and the diversification ratio. The variance 
explained by PC1 was calculated weekly using a rolling window size of 
two years (equivalent to 504 trading days). The diversification ratio was 
calculated using Equation \eqref{eqn:Diversification}.}
\label{fig:Diversification}
\end{figure}

The diversification ratio was discussed in Section (\ref{sub:Diversification})
above. Recall that it is a measure of the degree of diversification for 
a long-only portfolio.

Because the purpose of this paper is to study how the potential for 
diversification has changed over time with the systemic risk, not to 
compare how different allocation strategies result in different degrees of 
diversification, we only present the portfolio with the most simple 
allocation strategy, 1/N. Figure (\ref{fig:Diversification})
presents a plot of the diversification ratio for an equally weighted
portfolio of all 156 stocks together with the variance explained by
PC1.
 
The diversification ratio almost monotonically decreased with the increase 
in the variance explained by PC1. 
The diversification ratio dropped to its lowest point in our 
study period at approximately the same time the variance explained bu
PC1 reached its first peak.
This suggests that even if an investor was holding the same 
portfolio, it would have become less diversified leading into
the 2008 crisis.
 
Between the financial crisis in 2008 and the market decline in late 2011, 
diversification rose to about 3.75 at the same time
the variance explained by PC1 dropped a little 
but the diversification ratio 
was still 40\% lower than it was between 2002 and 2005. 
If we look at Figure (\ref{fig:pc1index}), we see that the index 
value recovered at the same time as the diversification ratio 
rose. 
Moreover, at the end of 2011, the diversification ratio dropped
when the variance explained by principal component one rose.
It is 
interesting that even at the end of 2011, the variance explained by 
principal component one rose to its second peak and was higher than it 
was in 2008, the diversification ratio, on the other hand, while 
historically low, was not lower 
than in 2008.

\section{Discussion and Conclusions}\label{sec:Discussion}

Above we have presented four methods of assessing the potential for
diversification in a single market. None of them answer the
criticism of \cite{meucci2009} in that they do not directly quantify
diversification potential. Nevertheless, as relative measures their
meaning is clear.  
Although they are not perfect
substitutes for each other they all show a consistent picture.
This is perhaps unsurprising given that they are all based in some way on
an analysis of correlation or covariance matrices. Of the
four, the simplest to
understand and easiest to use is the KMO test, which is typically 
applied to see if performing
a PCA is likely to result in a significant reduction in the
dimension of a data space. If we consider the return series for the 156 stocks
in our sample to be a 156 dimensional data space, if the data is
essentially a 156 dimensional hypersphere because each return
series is uncorrelated with the others, then the KMO statistic will be zero or
close to it, 
indicating a high potential for diversification. The further the data space
deviates from sphericity, intuitively this means it  
becomes elongated in one or more directions because
of common variation in returns, the higher the KMO statistic will be
hence indicating
less potential for diversification.

Sometimes this increase in common variation is referred to as
an increase in market connectedness. From the results above
we can see the formation of more connected market reduces the
potential to diversify a portfolio. 
Many researchers have reported that markets offer less diversification in a 
falling market \citep{ferreira2004,cappiello2006,Billio2012}. 
What our results show is that
when the Australian market was rising strongly the potential
for diversification was also decreasing. 
Thus if an investor held a 
constant-sized small portfolio between 2003 and 2008 that
portfolio became
less diversified over that period and offered little protection
when the crisis hit. 

The results of the PCA stock selection procedure in Figure 
(\ref{fig:nostocks0705}) initially appear counter-intuitive. 
Superficially it seems
to be recommending that an investor hold fewer stocks when the market
is more connected. It is, in fact, a more direct way to summarize the
diversification potential in the market. In the period 2002 to 2006
it tells us that the 156 stocks in our sample, which we are using
as a  proxy for the market, could be summarized with approximately
40 well-chosen stocks. By 2009 those same stocks could be summarized
in approximately 25 well-chosen stocks. Put in a more colloquial,
but readily understandable, manner, in the period 2002 to 2006 the
market had approximately 40 stocks' worth of diversification potential, by 2009
it only had 25 stocks' worth of diversification potential. Thus as the
diversification potential declined an investor who wished to hold a
well-diversified portfolio would need to seek
diversification opportunities in other asset classes to compensate
for the loss of such opportunities in the stock market.

Portfolio management is conducted in a larger context than simply finding
good stocks to buy.
Our investigation suggests that long before the financial crisis 
happened, the market had become more closely connected and that this
would have been fairly easy to detect. While all of the
methods we have presented suffer from the need to have an estimation
period all of them would have indicated that the ability to diversify
of a portfolio of stocks was declining well before it became a problem in the
2008 financial crisis. Both the diversification ratio and the
PCA portfolio selection method showed that
the potential to diversify within the market had decreased, that is,
adding more stocks would not have added much diversification to the
portfolio. The KMO statistic and the variance explained by PC1 rose indicating
an increase in common variation within the market.

Our results in 
the Australian market supports the observations in many papers such as
\cite{fenn2011}, \cite{Kritzman2011}, and \cite{Zheng2012} that systemic 
risk increased 
steadily in the years before 2008. We also found an increase of systemic risk
in Australian stock market around the end of 2011, which coincided with the 
European sovereign debt crisis. These two observations
are consistent with the study of 
systemic risk in the European market by \cite{Zheng2012}.
Our results are based on a similar testing framework to 
many other papers and adds two further supporting 
data points to the hypothesis that a large rise in the variance explained by PC1
may be a leading indicator of a financial crisis.


While the methods presented here were applied to a single stock market,
they are, in fact, quite general and can be applied to any set of 
investment opportunities for which a correlation matrix (or covarinace
matrix for the diversification ratio) can be 
generated.

\bibliography{Diversification}


\appendix

\section{Return Calculation }\label{App:Return} 

The return was calculated in the following steps:
\begin{enumerate}
\item We created a new variable associated with each stock
called the Dividend Factor. 
We started with a factor of 1 and every time a dividend was paid we 
multiplied the Dividend Factor, 

\begin{align*}
\text{Daily Dividend Factor}_{i}(t) & = \left\{ 
  \begin{array}{l l}
    1 & \quad \textrm{if no dividend}\\
    1+\frac{D_{i}(t)}{P_{i}(t)} & \quad \text{if dividend}
  \end{array} \right\} \\
\text{Cumulative Dividend Factor}_{i}(t)&= \prod_{j=1}^t (\text{Daily Dividend Factor}_{i}(t))
\end{align*}

where $D_{i}(t)$ is the dividend for stock $i$ in time $t$, $P_{i}(t)$ is 
price for stock $i$ at time $t$ in units of one trading day.
\item We adjusted the price series with the dividend factor, the adjusted 
price was calculated by
$$
\text{PNEW}_{i}(t)= P_{i}(t) \times \text{Cumulative Dividend Factor}_{i}(t).  
$$
\item The return series for a given stock $i$ was calculated as
\begin{equation}
 \text R_{i}(t)=\frac{\text{PNEW}_{i}(t+1)-\text{PNEW}_{i}(t)}{\text{PNEW}_{i}(t)}.
\label{eqn:returns}
\end{equation}
\end{enumerate}

\end{document}